\input{aipcheck}

\documentclass[
    ,final            % use final for the camera ready runs
  ]
  {aipproc}

\layoutstyle{8x11double}

\usepackage[usenames]{color}
\usepackage{epsfig}

\begin{document}

\title{New advances in photoionisation codes: How and what for?}

\classification{<Replace this text with PACS numbers; choose from this list:
                \texttt{http://www.aip.org/pacs/index.html}>}

\author{Barbara Ercolano }{
  address={Department of Physics and Astronomy, University College London, Gower Street, London WC1E 6BT, UK}
}

\begin{abstract}

The study of photoionised gas in planetary nebulae (PNe) has played a major role in the achievement, over the years, of a better understanding of a number of physical processes, pertinent to a broader range of fields than that of PNe studies, spanning from atomic physics to stellar evolution theories. Whilst empirical techniques are routinely employed for the analysis of the emission line spectra of these objects, the accurate interpretation of the observational data often requires the solution of a set of coupled equations, via the application of a photoionisation/plasma code. A number of large-scale codes have been developed since the late sixties, using various analytical or statistical techniques for the transfer of continuum radiation, mainly under the assumption of spherical symmetry and a few in 3D. These codes have been proved to be powerful and in many cases essential tools, but a clear idea of the underlying physical processes and assumptions is necessary in order to avoid reaching misleading conclusions. 

The development of the codes over the years has been driven by the observational constraints available, but also compromised by the available computer power. Modern codes are faster and more flexible, with the ultimate goal being the achievement of a description of the observations relying on the smallest number of parameters possible. In this light recent developments have been focused on the inclusion of new available atomic data, the inclusion of a realistic treatment for dust grains mixed in the ionised and photon dominated regions (PDRs) and the expansion of some codes to PDRs with the inclusion of chemical reaction networks. Furthermore the last few years have seen the development of fully 3D photoionisation codes based on the Monte Carlo method. 

A brief review of the field of photoionisation today is given here, with emphasis on the recent developments, including the expansion of the models to the 3D domain. Attention is given to the identification of new available observational constraints and how these can used to extract useful information from realistic models. 

\end{abstract}

\maketitle

%%%%%%%%%%%%%%%%%%%%%%%%%%%%%%%%%%%%%%%%%%%%
%% MAINMATTER
%%%%%%%%%%%%%%%%%%%%%%%%%%%%%%%%%%%%%%%%%%%%

\section{Introduction}

Photoionised plasma is present in many astrophysical environments, from H~{\sc ii} regions and Planetary Nebulae (PNe), that mark the beginning and the end stages of stellar evolution, to ionised interstellar and intergalactic media, to the gas photoionised by high energy sources in AGNs and seen in distant quasars. The interpretation of their complex emission line spectra relies on our ability to disentangle the underlying physics, which involves a number of microscopic atomic processes, highly sensitive to the physical properties of the emitting gas, and to the radiation field of the ionising source(s). Whilst empirical studies of the observations are routinely carried out to extract some basic information from the spectra, the application of large-scale numerical codes is often essential to the understanding of these sources. 

PNe, as many photoionised gas clouds, are generally transparent to the emission line radiation that is produced in their interior and, therefore, a prediction of their spectra simply requires a knowledge of the electron temperature and ionisation structure at each position in the ionised volume. This can be achieved by numerically solving the coupled ionisation balance equations and the thermal equilibrium. The local radiation field, including the stellar and diffuse components, must first be estimated by, classically, analytical or, more recently, statistical means. All major heating and cooling channels must be accounted for as well as all ionisation and recombination processes. A comprehensive reference is the textbook by Osterbrock (1989), additionally, a review on quantitative spectroscopy of photoionised clouds was given by Ferland (2003), whilst a summary of emission line analysis can be found in a number of papers presented at a conference to honour Silvia Torres-Peimbert and Manuel Peimbert on their sixtieth birthdays (Henney et al. 2002) and those presented at a previous meeting organised to celebrate the seventieth birthday of Don Osterbrock and Mike Seaton (Williams \& Livio 1995).

The first generation of codes that were able to treat the temperature stratification in H~{\sc ii} regions appeared in the late 1960s (e.g. Flower 1968, Rubin 1968, Harrington 1968), these codes were severely hindered by the limited computing power available at the time and the lack of a comprehensive atomic data set. The fast improvements in hardware and the large volume of atomic data become available in the last forty years have allowed photoionisation calculations to become more and more sophisticated boosting the predicting power of the codes and making them an essential tool for modern quantitative spectroscopic studies. At present  more than eight large scale photoionisation codes are in use. Their performance has been tested through a number of workshops, resulting in the definition of a set of rigorous benchmarks (P\'{e}quignot et al. 2001), generally referred to as the Meudon/Lexington benchmarks. The state of the art of the field is accurately described by a number of papers presented at the latest of these meetings and included in {\it Spectroscopic Challenges of Photoionised Plasmas} (Ferland \& Savin 2001). In the present paper I summarise where photoionisation modelling is today, highlighting the major developments that have occurred in the past five years, with particular emphasis on the applications of modern plasma codes to the modelling of PNe.

\section{New advances: How?}

%!\begin{figure}[t]
%!  \includegraphics[width=7.cm, height=.3\textheight]{fig1.eps}
%!  \caption{Scheme of photoevaporation flow from Henney et al. 2005a}
%!\label{fig:1}
%!
%\end{figure}

The construction of a photoionisation model aims to match all available constraints for a given target, which may include spatially resolved or not, optical, infrared (IR) and/or ultraviolet (UV) spectra. Images in one or more narrow-band filter are also available for many objects. A well-constrained tailored photoionisation model can unveil a large volume of information regarding the nebular properties, including the gas density distribution and its chemical composition, and regarding the properties of the ionising star(s), such as effective temperature and surface gravity. Additionally, models of PNe have been used in the past to constrain the magnitude of some then unknown atomic process rates, as in the case of the hydrogen-oxygen charge exchange process (P\'equignot, Stasinska \& Aldrovandi, 1978). 

The desire of obtaining {\it realistic} models for a variety of emission line sources is the main drive behind the continuous developments of large photoionisation codes in the last forty years, including the more recent advances that will be discussed here. Progress has been made in many aspects, in this paper the follow five main streams will be considered: atomic data updates, the treatment of time-dependent effects, the inclusion of dust grains, the expansion of the models to the PDR and, finally, the development of 3-D codes. These will be discussed in more detail in the remainder of this section. 

\subsection{Atomic data updates}

\begin{figure*}

\epsfxsize=0.3\textwidth\epsfbox[120 318 450 653]{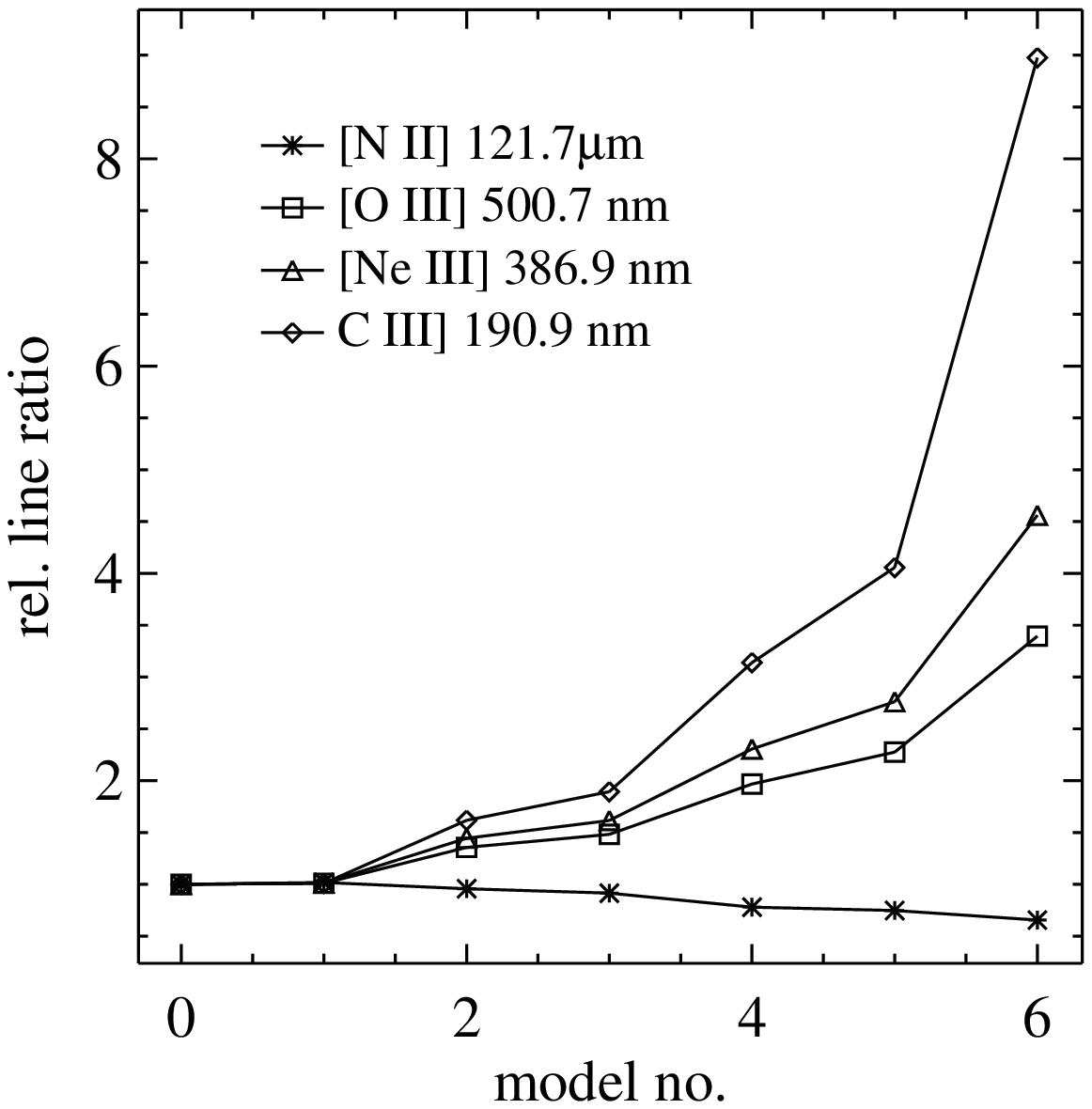}\hfill
\epsfxsize=0.3\textwidth\epsfbox[120 318 450 653]{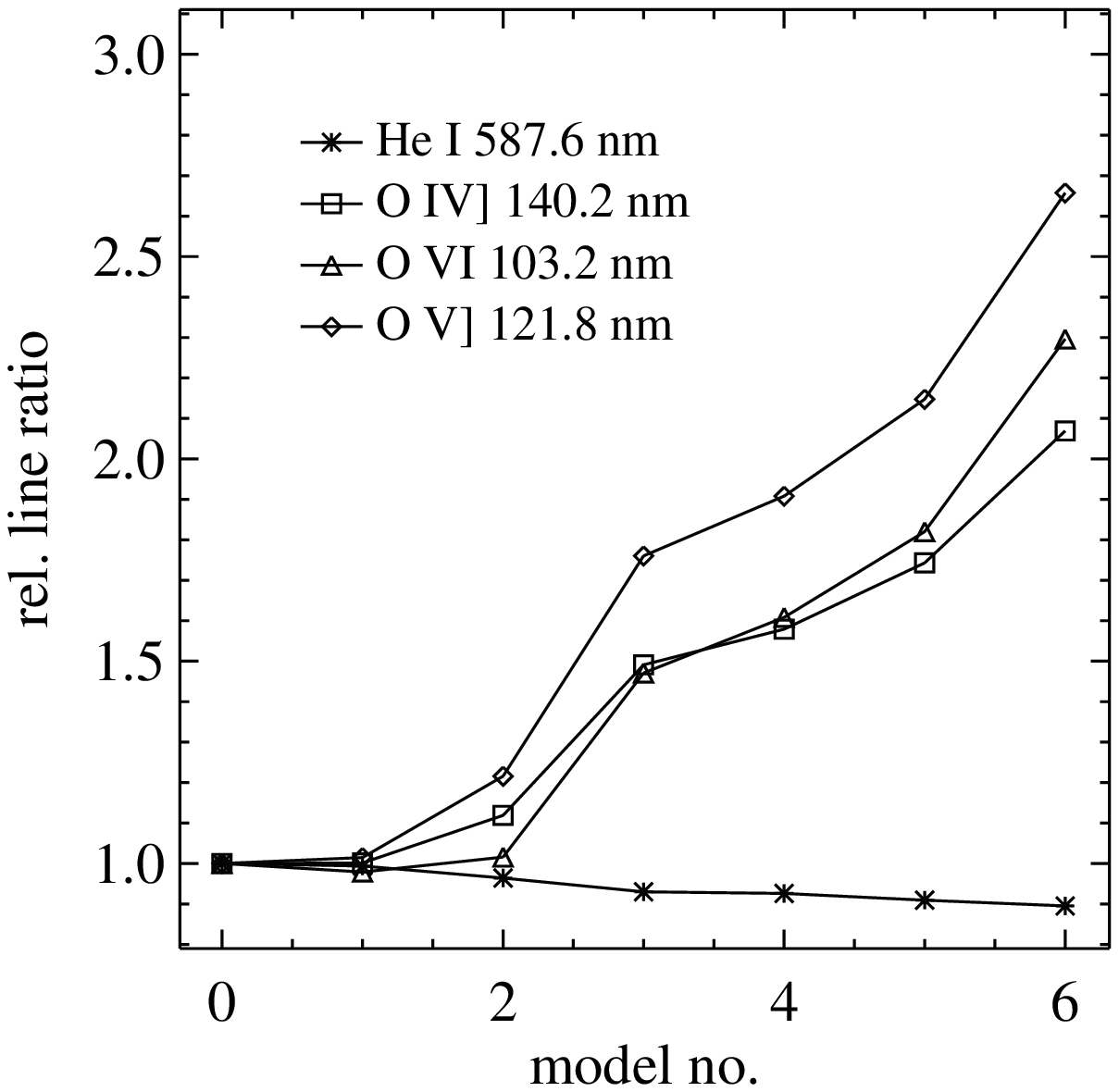}\hfill
\epsfxsize=0.3\textwidth\epsfbox[120 318 450 653]{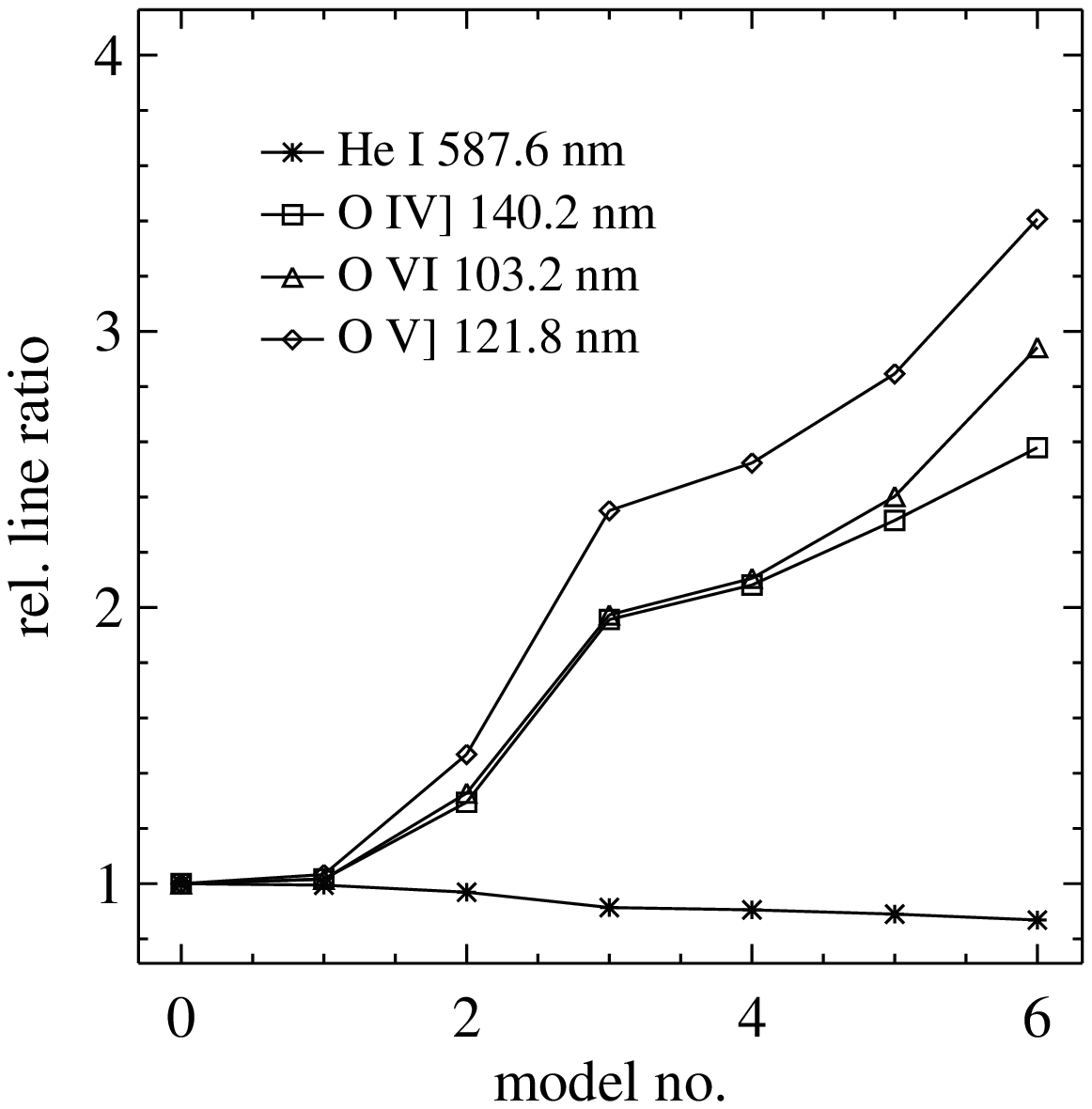}\hfill

\caption{(van Hoof et al. 2004) The variation of the line ratio to H$\beta$ for selected emission
lines as a function of the grain size distribution. The left panel shows the
results for the H\,{\sc ii}-region models, the middle panel for the PN models
with silicate, and the right panel for the PN models with graphite. All line
flux ratios are normalised to the base model without dust. The models numbers
are defined in Table~4 of van Hoof et al. (2004).}
\end{figure*}

Advances in computer hardware and the construction of
large scale theoretical atomic codes that can compute
photoionisation cross-sections, electron-ion recombination rates,
level energies, electron-collision strengths, transition probabilities
and oscillator strengths for a large number of ions and transitions 
have allowed a comprehensive data set
to be compiled. In particular, the Opacity Project and subsequently
 the Iron Project (Berrington et al. 1987), have produced a wealth 
of astrophysically useful atomic data, that is readily accessible via 
the TIPTOPbase website (see e.g. Mendoza et al. 2002, Nahar 2003).

New total recombination coefficients, which include both radiative and
dielectronic contributions, have recently been presented for a number
of ions (e.g. Nahar 2005, and references therein). These are ab-initio
calculations that simultaneously and self-consistently treat
photoionisation and recombination processes, although in some cases
they rely on theoretical predictions of the resonance positions. The
emphasis has recently been on higher ionisation
species, not expected in PNe. New calculations of total recombination
coefficients of some third-row elements in ionisation stages commonly
observed in PNe are currently under way (Storey, Ercolano \& Badnell,
in preparation). These use the {\sc autostructure}
program of Badnell (see e.g. Badnell 1999)
in the LS- or intermediate-coupling (IC) schemes, where
necessary, with corrections to the theoretical level energies being
applied, where experimental values are known. The coefficients are
calculated for a wide temperature range and can also be applied to very
cool (approx. 500~K) plasma.

Observations of optical recombination lines (ORL) in long-slit and
integral field unit (IFU) spectra are nowadays readily obtainable. The
analysis of such spectra poses some of the most
intriguing questions yet to be answered in nebular astrophysics,
namely the fact that ionic abundances derived from
ORLs are consistently higher than those derived from the CELs in the
same spectra, whilst the former diagnose temperatures that are
significantly lower than those estimated using CEL ratios, He~{\sc i}
recombination line ratios or the Balmer decrement
(see e.g. Rubin 1997, Liu et al. 2000, Liu et al. 2001b, Tsamis et al. 2004). 
Photoionisation models of clouds containing chemical
inhomogeneities and/or density discontinuities may be crucial to clarify
this long-standing problem (e.g. P\'{e}quignot et al. 2003, Tylenda 2003,
Ercolano et al. 2003), however they rely heavily on accurate
recombination data extending to low-temperatures. Work is currently
being carried out at UCL for a number of ions and some preliminary
results have been presented by Bastin \& Storey (2005a,b these
proceedings).

\subsection{Time-dependence effects}
\begin{figure}[t]
  \epsfig{file=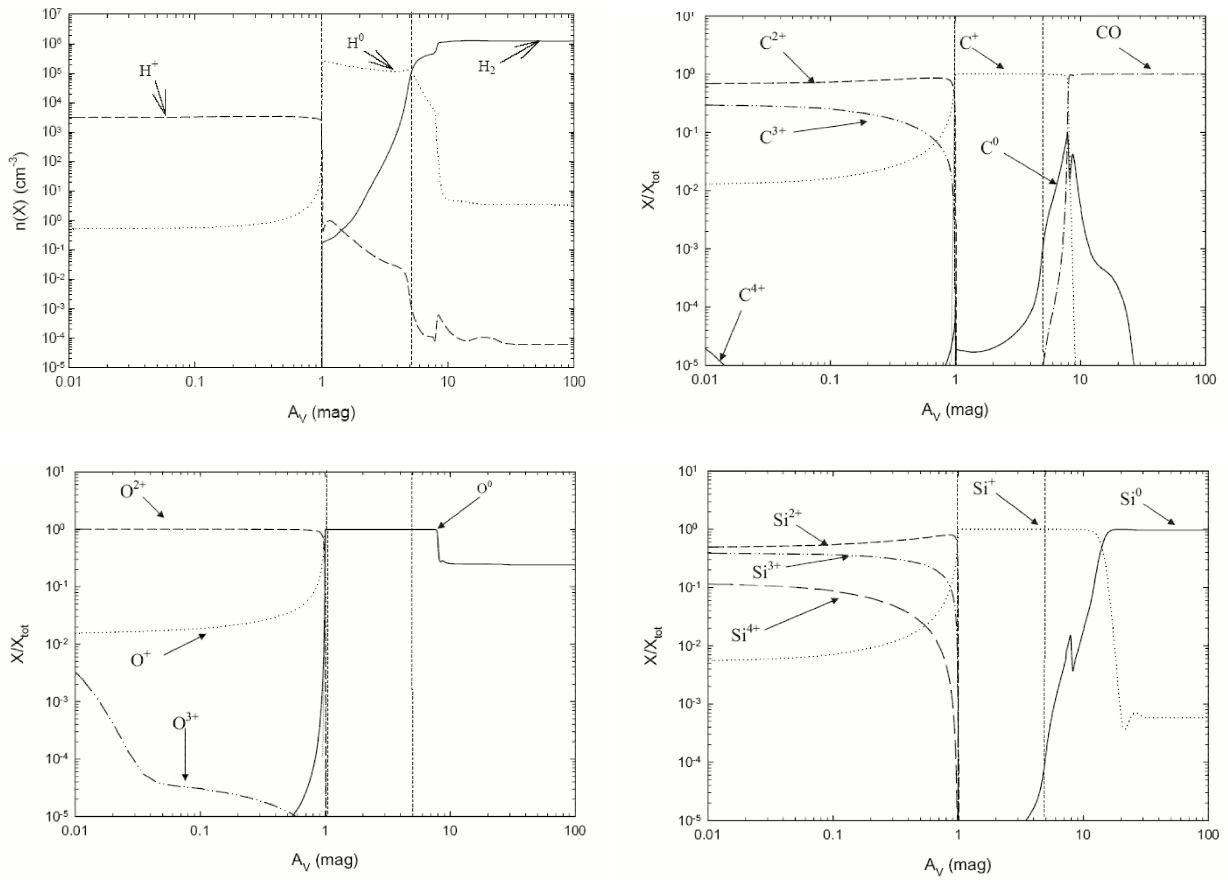,width=15.5cm,clip=,bbllx=114, bburx=480,bblly=300,bbury=561}
\caption{Ionisation structure of a standard cloud from Abel et al. 2005; the two vertical  lines represent the hydrogen ionisation front and the half-H$_2$ point. {\it Top left} The H$^+$,H$^0$ and H$_2$ density structure. {\it Top right} The C ionisation structure. {\it Bottom left} The O ionisation structure. {\it Bottom right} The Si ionisation structure.}
\label{fig:abel05}
\end{figure}

The ionisation structure and electron temperature of a photoionised cloud is obtained by simultaneously solving the thermal balance and ionisation equilibrium equations, classically under the steady-state assumption. This approximation is valid when the atomic physics timescales are short compared to those of gas-motion or to the rate of change of the ionising field. If this is true the photoionisation problem can be considered to be fully decoupled from the dynamics; in this case a given gas density distribution represents a {\it snapshot} of the cloud evolution and a time-independent photoionisation simulation can be performed. PNe and H~{\sc ii} can generally be expected to be mostly in equilibrium, with dynamical effects only becoming significant in a very small region near the ionisation front, hence having near negligible effects on the integrated emission line spectrum (e.g. Harrington, 1977).  

The cases studied by Harrington (1977) were limited to weak D-fronts with subsonic gas motions with respect to the front. There are, however, a number of instances when dynamics may have substantial effects on the predicted spectrum. Transonic photo-evaporation flows, such as those believed to be occurring in cometary knots of PNe (L\'{o}pez-Mart\'{i}n et al. 2001), cannot be accounted for under the assumption of thermal and ionisation equilibrium. 

In the cases when shocks are present the gas cannot be considered to be in equilibrium, and classical time-steady codes are not appropriate. The {\sc mappings~{\sc iii}} code (Sutherland \& Dopita 1993) was designed to deal with shock ionisation and has been applied to date to a variety of astrophysical environments. This code have recently undergone substantial updating and it is publicly available on-line via an interactive web interface (see http://cfa-www.harvard.edu/~lkewley/Mappings/).

Time-steady dynamics have recently been added to the widely used plasma code {\sc cloudy} (publicly available at http://www.nublado.org/); the numerical algorithms employed and a set of self-consistent dynamic models of steady ionisation fronts are presented in recent papers by Henney et al., 2005a,b. 

\subsection{Treatment of dust grains}

The importance of the effects of dust grains that are mixed with the gas in a photoionised region has long been known (e.g. Spitzer 1948), with more recent studies also aimed at investigating the role of dust grains in the thermal balance of PNe (e.g. Borkowski \& Harrington 1991, Ercolano et al. 2003c, van Hoof et al. 2004). The grains are heated via the absorption of UV photons from the continuum as well as resonance emission line photons, for example H~{\sc i} Ly$\alpha$,  C~{\sc iv}\,$\lambda$1549, N~{\sc v}\,$\lambda$1240, C~{\sc ii}\,$\lambda$1336, Si~{\sc iv}\,$\lambda$1397, Mg~{\sc ii}\,$\lambda$2800. The absorbed radiation is mainly re-emitted by the grains in the IR, and this provides the main dust cooling mechanism. Additionally, the gas and dust components are coupled by a host of microphysical processes, including photoelectric emission from dust grains (for a recent discussion see Weingartner \& Draine 2001), which may be an important gas-heating mechanism in H-deficient environments and in PDRs. Gas-grain collisions provide an extra cooling channel for the gas and a heating channel for the dust, and ionic recombination on dust grains may also have a small effect on the ionisation structure of the gas. 

 Over the last thirty years, a number of photoionisation codes have included some kind of treatment to allow for dust opacities to be taken into account for in the ionised region and/or obtain estimates for the dust photoelectric heating. This topic, however, has recently been revisited by some authors in view of the expansion of existing codes to allow the construction of self-consistent models of ionised regions with their surrounding PDRs, where dust provides the main source of opacity. This will be be discussed in more detail in the next section. An example is the new grain model that has recently been developed for {\sc cloudy} (van Hoof et al. 2004). The major updates include a more flexible structure, allowing for grain sizes and species to be treated individually, rather than using a {\it mean} grain to be representative of an ensemble. This is important if the contribution of small grains to the SED is to be accurately predicted. Furthermore, these authors presented a more sophisticated treatment of the grain charge problem, inspired by the work of Weingartner \& Draine (2001); the {\it hybrid} grain model improves on the previously implemented {\it average grain charge}  model (Baldwin et al. 1991), allowing a more accurate calculation of photoelectric yields. Figure~2 (van Hoof et al. 2004) shows the effects of dust on the emission line spectra of typical PNe and H~{\sc ii} regions models, obtained by including a dust component to the Meudon/Lexington benchmarks. The models in the plots show some dramatic departures from the dust-free case; it should be noted, however, that these models extend to the PDR region, while the effects on the photoionised region alone are expected to be smaller in magnitude. 

A new version (2.0) of the 3D Monte Carlo photoionisation code {\sc mocassin} has recently been released (Ercolano et al. 2005), which includes a fully self-consistent treatment of the dust radiative transfer within the ionised region. Dust grains are allowed to compete with the ions in the gas for the absorption of UV and resonance emission line radiation, with scattering by the grains being fully implemented in the isotropic approximation or via the application of a phase function. The microphysical processes listed above, which further couple the two components, are also accounted for in the thermal and ionisation balance equations. The average grain charge model was preferred by these authors to a fully quantised or hybrid treatment in order to limit overheads, in light of the large uncertainties still existing in the dust data (see Weingartner \& Draine 2001, Sec. 2.3). The {\sc mocassin} code, which can also be run in a pure-dust mode, has the advantage of being fully three-dimensional, hence allowing the scattering problem to be treated. Work is currently in progress to allow polarisation maps to be produced, which may yield better constraints on the geometry of the many sources for which this type of observations are available. 

\subsection{Expansion to Photon Dominated Regions}

Photon Dominated Regions (PDRs -- also known as {\it photo-dissociation regions}) are associated with all photoionised nebulae, including PNe, that are optically thick to 13.6~eV radiation. When all photons with $h\nu\,>\,$13.6\,eV have been absorbed H becomes predominantly neutral, in general marking the beginning of the PDR (Tielens \& Hollenbach, 1985). Classically, the two regions have been treated as separate problems, both in terms of empirical analyses based on spectroscopic observations and in terms of the development of numerical codes that would be able to model either of the two environments. In reality, a photoionised region and its PDR are closely coupled and, therefore, a self-consistent calculation is preferable. The radiation field impinging on the edge of a PDR is, in fact, a result of photon diffusion through the ionised region, and is naturally calculated by a standard photoionisation code, whilst classical PDR codes generally treat G$_0$\footnote{$G_0$ is defined as the Far Ultra Violet (FUV) field between 6 and 13.6~eV relative to the background interstellar radiation field (Habing, 1968)} as a free parameter to be determined from the model. Moreover, some of the most important IR diagnostics used to characterise the gas in a PDR are also partially emitted in the ionised region. This is the case, for example, for all transitions of ions such as O$^0$, C$^+$, Si$^+$, having Ionisation Potentials (IPs) lower than 13.6~eV. Figure~\ref{fig:abel05} shows a plot of the ionisation structure of an H~{\sc ii} region and its associated PDR, as recently calculated by Abel et al. (2005). 

Many PNe are known to be surrounded by extended PDRs, as is the case, for example, of NGC\,7027 (e.g. Graham et al. 1992, Latter et al. 2000) and NGC\,6302 (e.g. Liu et al. 2001a). More recently, PDRs have also been detected around cometary knots, such as those of the Helix nebula (O'Dell, Henney \& Ferland 2005) and the Ring nebula (Speck et al. 2003). 

The {\sc cloudy} team are again leading the way, with a new version of their code (C05.07.06) recently released, which allows self-consistent calculations of the spectrum, chemistry, and structure of photoionised clouds and their associated PDRs to be performed (Abel et al. 2005). Besides the improvements to the grain model mentioned in the previous section, a molecular network was also included (Abel et al. 2005) with approximately 1000 reactions involving 68 molecules. Recent advances in the treatment of H$_2$ are described by Shaw et al. (2005). The chemical equilibrium was benchmarked at the Leiden 2004 PDR workshop (http://hera.ph1.uni-koeln.de/$\sim$roellig/, Roellig et al. in prep.) and found to be in agreement with the results of other PDR codes.  

Having completed the inclusion of dust grains and subsequent benchmarking phase, the {\sc mocassin} code is also currently being expanded to incorporate the molecular reaction network in use by the {\sc ucl\_pdr} code (Bell et al. 2002, and references therein), which is amongst the codes recently benchmarked at the Leiden 2004 PDR workshop. The new version of the {\sc mocassin} code, which will be described in a forthcoming paper by Ercolano, Bell, Barlow \& Viti., will allow one to make self-consistent models of ionised regions and PDRs of arbitrary geometry and density distributions, including, for example, cometary knots and their shadow regions. 

\subsection{Development of 3-D codes}

Available computer power has increased enormously since the development of the first generation of photoionisation codes in the late sixties. This has allowed the construction of more complex models, that, thanks also to the great advances in atomic physics, may include more ions, more frequency points, more lines and more atomic levels. Nevertheless, with a few exceptions, the fundamental assumption of spherical symmetry has always been retained. A glance at spatially resolved images of H~{\sc ii} regions and PNe, that are readily obtainable with modern instrumentation, immediately demonstrates that these object are rarely circular in projection. The effects of geometry and/or density distribution on the emerging emission line spectrum are in some cases dramatic, and an unrealistic model may result in the incorrect determination of nebular properties (e.g. elemental abundances) and/or central star parameters, which may yield misleading conclusions. 

The careful combination of a number of volume-weighted spherically symmetric models (composite models) has been used to treat some aspherical geometries, including bipolar structures (e.g. Clegg et al. 1987) or density/chemical enhancements (P\'{e}quignot 2003). Morisset et al. (2005) have recently presented a pseudo-3D photoionisation code, based on this principle. The code consists of an $IDL (RSI)$ suite of programs that make several calls to a 1D code for a number of selected lines of sight through a 3D structure. The 1D code currently used is the {\sc nebu} code (Morisset \& P\'equignot 1996, P\'equignot et al. 2001), but the technique can be readily applied to any other 1D photoionisation code. Whilst this method has the advantage of being considerably faster than running a full 3D simulation, one of its major drawbacks is that the transfer of the diffuse component of the radiation field cannot, in general, be treated self-consistently. Ercolano et al. (2003b) found appreciable discrepancies between the results obtained by a self-consistent calculation of even a simple biconical geometry and those obtained by a composite model. 

To my knowledge the first two 3D photoionisation codes were those developed by Baessgen et al. (1990) and by Gruenwald et al. (1997). The latter is still in use and 3D models of aspherical PNe obtained with this code have recently been published (Monteiro et al. 2005, Monteiro et al. 2004). Both codes used analytical approaches to the transfer of radiation; the first treats the diffuse field in the on-the-spot approximation, whilst the second employs an iterative technique for the determination of the diffuse field.

The fast development of parallel computing seen in last few year, has made the application of stochastic techniques to the radiation transport problem an attractive option. The first 1D Monte Carlo photoionisation calculations were those of Och et al. (1998). The first fully 3D Monte Carlo photoionisation code, {\sc mocassin} (Ercolano et al. 2003a), uses a similar approach, but a more general version of the Monte Carlo estimators for the radiation field, developed by Lucy (1999). The description of the radiation field in terms of discrete quantities allows one to simulate the individual scattering, absorption and re-emission events that characterise a photon's trajectory as it diffuses through a cloud. This public code is fully parallelised and looks to exploit the increasing accessibility of large Beowulf clusters to allow the construction of realistic nebular models. As mentioned above, the inclusion of grain physics has been one of the major recent updates of the code (Ercolano et al., 2005), in preparation for the expansion to PDR, planned for the near future. 

Wood, Mathis \& Ercolano (2004) also presented a 3D Monte Carlo photoionisation code, based on similar techniques as those described by Och et al. (1998) and Lucy (1999). This code is tailored for the modelling of Galactic H~{\sc ii} regions and the percolation of ionising photons in diffuse ionised gas (Wood \& Mathis, 2004) .

\subsection{3-D photoionisation models of PNe}

\begin{figure*}
\begin{minipage}{7.051cm}
  \epsfig{file=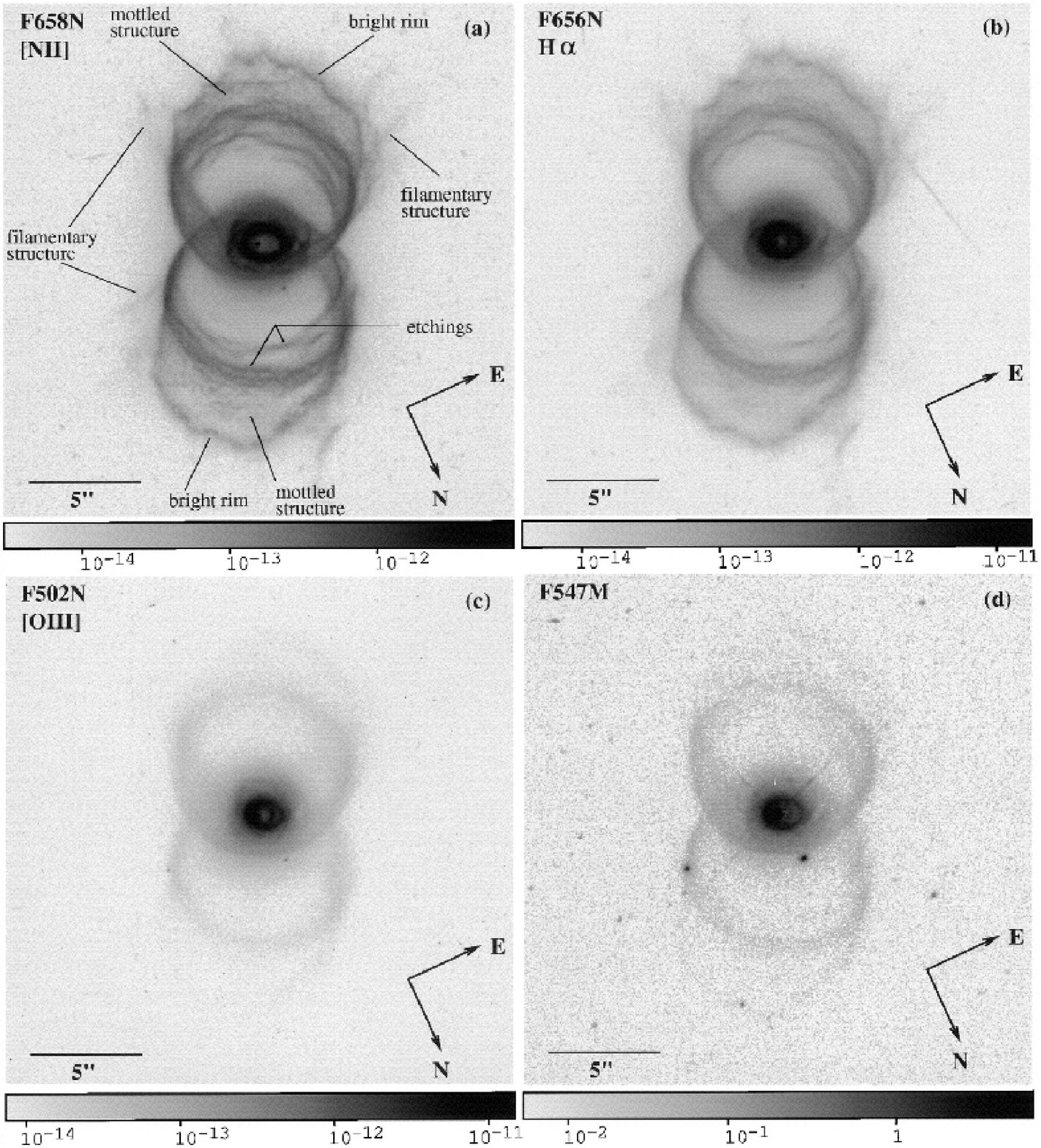,width=7.051cm}
\end{minipage}
\begin{minipage}{7.051cm}
  \epsfig{file=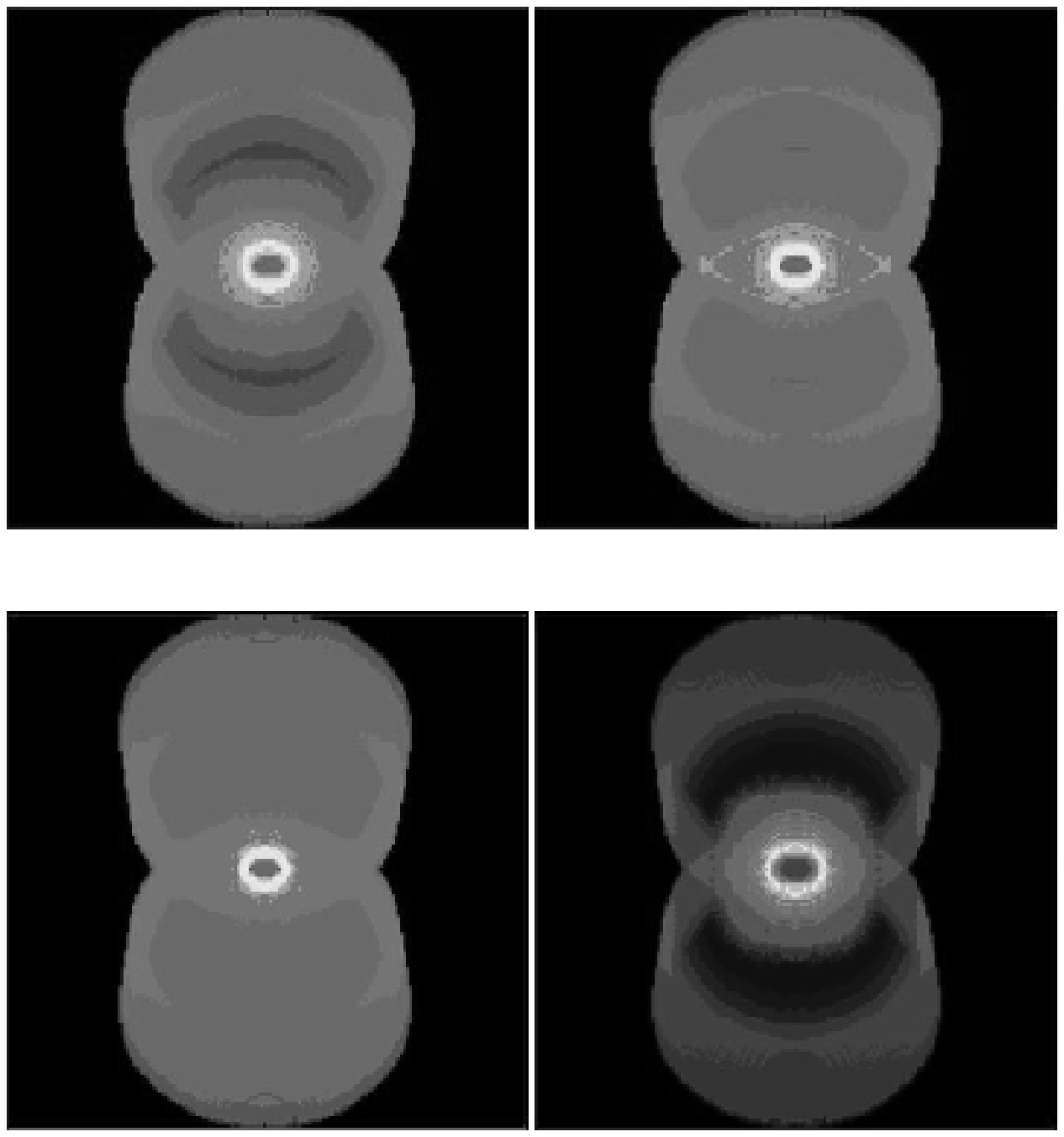,width=7.051cm,height=7.7cm,clip=,bbllx=110, bburx=479,bblly=218,bbury=600}
\end{minipage}

\caption{\emph{Left Panel}: From Sahai et al. 1999. Narrow-band HST WFPC2 images of MyCn 18: (a) F658N ($[$N~{\sc ii}$]\lambda$6586; (b) F656N (H$\alpha$); (c) F502N ($[$O~{\sc iii}$]\lambda$5007); (d) a continuum filter F547N. \emph{Right panel}: From Neal et al. (in prep.). 2D projections of 3D emissivity grids obtained from our best-fitting {\sc mocassin} model of MyCn18: (a) $[$N~{\sc ii}$]\lambda$6586; (b) H$\beta$); (c) $[$O~{\sc iii}$]\lambda$5007; (d) $[$O~{\sc i}$]\lambda$6300.}
\label{fig:mycn18}
\end{figure*}

In this section some recent results of 3D pure (no dust) photoionisation calculations recently obtained with the {\sc mocassin} code are summarised. 

Figure~\ref{fig:mycn18} (Neal, Ercolano \& Sahai, in preparation) show the projected maps obtained from a model of the young PN MyCn18, compared with the {\it Hubble} images presented by Sahai et al. (1999). The model images are shown to be in very good agreement with  the observations, indicating that the model 3D density distribution for this object, previously presented by Dayal et al. (2000) and employed in our simulation is a realistic description for this object. A well constrained density distribution adds confidence to the model predictions of the other nebular and stellar physical properties.   

\begin{table}
\begin{tabular}{lrr}
\hline
   \tablehead{1}{l}{b}{Line Identification ({\AA})}
  & \tablehead{1}{c}{b}{No Slit}
  & \tablehead{1}{c}{b}{Slit}  \\
\hline
He~{\sc i} 5876          & 14.1 & 13.4  \\
He~{\sc ii} 4686         & 16.8 & 22.0  \\
$[$O~{\sc ii}$]$~3726,29 & 5.38 & 5.88  \\
$[$O~{\sc iii}$]$~5007   & 1197 & 1172  \\
$[$N~{\sc ii}$]$~6584    & 5.87 & 6.64  \\
$[$S~{\sc ii}$]$~6731    & 0.84 & 0.99  \\
$[$S~{\sc iii}$]$~6312   & 1.47 & 1.55  \\
\hline
\end{tabular}
\caption{Model predicted emission lines for NGC~7009. Subset of model spectrum obtained by Gon\c calves, Ercolano et al., 2005a.}
\label{tab:n7009}
\end{table}

The effects of observations with a finite aperture on non-spherically symmetric objects should also be taken into account, their magnitude being larger the greater the departure from spherical symmetry of the observed geometry. As an example, long-slit spectroscopic observations of extended objects, such as H~{\sc ii} regions and PNe, are routinely performed. Whilst line ratio measurements obtained from the summed spectra of a long-slit scanned across the whole 2D projection of a nebula may be considered to be representative of the integrated nebular spectrum, they are not always available. Geometry effects on long-slit spectra at a single PA were recently studied in the case of the PN NGC~7009 (Gon\c calves, Ercolano et al., 2005a, Gon\c calves et al., 2005b, these proceedings) and were found to be partially responsible for the long-standing apparent overabundance of N in the FLIERs of this object (Balick et al.~1994, Hajian et al.~1997, Balick et al.~1998, and references therein). Table~\ref{tab:n7009} lists a subsection of the slit-less (column~2) and slit (column~3) {\sc mocassin} model emission line spectra of the FLIERs of NGC~7009, showing an increase of approximately 13\% in the predicted flux of $[$N~{\sc ii}$]$~$\lambda$6584. These results, discussed in more detail by Gon\c calves, Ercolano et al., 2005a, show that care should be taken when applying ICF methods or unrealistic photoionisation models to non-spherically symmetric objects. In this light, even the classically accepted type~{\sc i}/{\sc ii} classification scheme of PNe may become questionable. 

Chemical inhomogeneous models of the hydrogen-deficient polar knots of the PN~Abel~30 were presented by Ercolano et al., 2003c; these models showed that it is in fact possible to explain the extreme ORL-CEL abundance discrepancy factors (ADFs) reported for this object by a simple bi-abundance model, consisting of a dense metal-rich (cold) core of ionised gas mixed with dust grains surrounded by a less dense envelope of hydrogen deficient gas with less extreme metal abundances. In this model, ORLs of heavy elements result are shown to be emitted almost exclusively by the cold core, CELs from the hotter surrounding envelope, whilst He recombination lines are emitted by both components. 

In the case of some H~{\sc ii} regions for which only mild ADFs are observed, the discrepancy may be resolved by considering temperature fluctuations in the t$^2$ prescription of Peimbert (1967). Whilst this interpretation has been successful in explaining the observations for some objects with mild ADFs, the nature and cause of the supposed temperature fluctuations remain uncertain. Even models that include density fluctuations yield only small deviations from an otherwise near homogeneous temperature structure of the emitting regions. One possibility worth investigating is whether larger fluctuations may be produced with a model including multiple ionising stars. A number of 3D {\sc mocassin} models are being constructed which use several geometries and ionising source distributions (Ercolano, Stasi\'nska \& Barlow, in preparation), which may help to clarify the situation from a theoretical point of view. 

\section{Conclusion: The Future}

It has been the aim of this review to identify and discuss in broad terms the most recent advances in photoionisation codes. Several large scale codes exist and are being maintained by individual working groups, coming together regularly to ensure the reliability of the results via the performance of a set of benchmark problems. Modern codes aim at the construction of realistic simulations in order to efficiently extract information from a wealth of new spectroscopic observations available at large wavelength ranges and for increasingly more distant and faint objects. 

The coupling of photoionisation/PDR and hydrodynamic RT codes is only a step away. This is perhaps one of the most exciting prospects, as the new simulations will undoubtedly yield new insights into the physics of many astronomical phenomena. Already some photoionisation codes allow for time-dependent effects to be taken into account. In the near future it will be possible to perform a full photoionisation/PDR calculation at each time-step in a large hydrodynamic simulation, to follow, for example, the evolution of the time-dependent non-equilibrium spectrum of a dynamically evolving gas cloud, from the time it starts being illuminated. Both physics and numerical techniques are mature, it is only a matter of waiting for the computer power required to become available. 

\begin{theacknowledgments}
I would like to thank the SOC \& LOC for inviting me to give this review talk and organising a very stimulating meeting. 

\end{theacknowledgments}



\bibliographystyle{aipprocl} % if natbib is missing

\bibliography{sample}

\begin{thebibliography}{9}

\bibitem{a} N. P. Abel, G. J. Ferland, G. Shaw, P. A. M. van Hoof, 2005, ApJ in press

\bibitem{b} N. R. Badnell 1999, J.Phys.B 32 p. 5583

\bibitem{c}	
M. Baessgen, C. Diesch, M. Grewing, A\&A, 1990, A\&A 237, 201

\bibitem{d}
B. Balick, J. Alexander, A. R. Hajian, Y. Terzian
 \& M. Perinotto, 1998, AJ 116, 360

\bibitem{e}
B. Balick, M. Perinotto, A. Maccioni, J. Alexander, Y. Terzian
 \& A. R. Hajian, 1994, ApJ 424, 800

\bibitem{ea}
T. A. Bell, S. Viti, D. A. Williams, I. A. Crawford, R. J. Price, 2005, MNRAS 357, 961

\bibitem{f}
R. Bastin \& P. J. Storey, 2005a, these procs.

\bibitem{g}
R. Bastin \& P. J. Storey, 2005b, these procs.

\bibitem{h}
K. J. Borkowski \& P. J. Harrington 1991, ApJ 379, 168 

\bibitem{i}
K. A. Berrington, P. G. Burke et al., 1987, J.Phys.B 20, 6379

\bibitem{j}
 R. E. S. Clegg, J. P. Harrington, M. J. Barlow, J. R. Walsh, 1987, ApJ 314, 551

\bibitem{k}
A. Dayal, R. Sahai, et al., AJ 119, 315

\bibitem{l}
B. Ercolano, M. J. Barlow, P. J.  Storey,  X.-W. Liu, 2003a, MNRAS 340, 1136

\bibitem{m}
B. Ercolano, C. Morisset, M. J. Barlow, P. J.  Storey,  X.-W. Liu, 2003b, MNRAS 340, 1153

\bibitem{n}
B. Ercolano, M. J. Barlow, P. J.  Storey,  X.-W. Liu, T. Rauch, K. Werner, 2003c, MNRAS 344, 1145

\bibitem{o}
B. Ercolano, M. J. Barlow \& P. J.  Storey, 2005, MNRAS, in press, astro-ph/0507050

\bibitem{p}
B. Ercolano, G. Stasi\'nska \& M. J. Barlow, MNRAS, in preparation

\bibitem{q}
B. Ercolano, T. A. Bell, M. J. Barlow,  S. Viti \& P. J.  Storey, MNRAS, in preparation

\bibitem{r}
G.~J. Ferland, D.~W. Savin, eds 2001. \emph{Spectroscopic Challanges of Photoionised Plasmas}, ASP Conf. Ser. 247. San Francisco: Astron. Soc. Pac.

\bibitem{s}
G.~J. Ferland, 2003, ARA\&A 41, 517

\bibitem{t}
D.~R. Flower, 1968, ApJ 2, L205

\bibitem{u}
D. Gon\c calves, B. Ercolano, A. Carnero, A. Mampaso \& R. Corradi, 2005a, MNRAS, in press

\bibitem{v}
D. Gon\c calves, B. Ercolano, A. Carnero, A. Mampaso \& R. Corradi, 2005b, these proceedings

\bibitem{w}
 J. R. Graham, E. Serabyn, T. M. Herbst, K. Matthews, G. Neugebauer, B. T. Soifer, T. D. Wilson, S. Beckwith 1992, AAS 181, 3903

\bibitem{wa}
T. R. Roellig et al., A\&A, in prep.

\bibitem{x}
R. Gruenwald, S. M. Viegas, D. Broguiere, 1997, ApJ 480, 283

\bibitem{y}
H. J. Habing, 1968, Bull. of the Astron. Inst. of the Neth. 19 , 421

\bibitem{z}
A. R. Hajian, B. Balick, Y. Terzian \& M. Perinotto M., 1997, ApJ 304, 313

\bibitem{za}
P.~J. Harrington, 1977, MNRAS 179, 63

\bibitem{zb}
P.~J. Harrington, 1968, ApJ 152, 943

\bibitem{zc}
W.~J. Henney, J. Franco, M. Martos, M. Pe\~{n}a, eds. 2002. \emph{Ionized Gaseous Nebulae, a Conference to Celebrate the 60th Birthdays of Silvia Torres-Peimbert and Manuel Peimbert}. Rev. Mex. Astron. Astrof\'is. 12

\bibitem{zd}
W.~J. Henney, S. J. Arthur, R. J. R. Williams, G. J. Ferland, 2005a, ApJ 621, 328

\bibitem{ze}
W.~J. Henney, S. J. Arthur, Ma. T. Garci-Diaz, 2005b, ApJ 621, 328

\bibitem{zf}
W. B. Latter, A. Dayal, J. H. Bieging, C. Meakin, J. L. Hora, D. M.  Kelly, A. G. G. M. Tielens, 2000, ApJ 539, 783

\bibitem{zg}
X.-W. Liu, M. J.  Barlow, M. Cohen, I. J. Danziger, S.-G. Luo, J. P. Baluteau, P. Cox, R. J. Emery, T. Lim, D. Péquignot, 2001a, MNRAS 323, 343

\bibitem{zh}
X.-W. Liu et al., 2001b, MRAS 323, 343

\bibitem{zi}
X.-W. Liu, P. J. Storey,  M. J. Barlow, I. J. Danziger, M. Cohen, M. Bryce, 2000, MNRAS 312, 585

\bibitem{zl}
I. L\'{o}pez-Mart\'{i}n, A. C. Raga, G. Mellema, W. J. Henney \& J. Cant\'{o}, 2001, ApJ 548, 288 

\bibitem{zm}
H. Monteiro, H. E. Schwarz, R. Gruenwald, K. Guenthner, S. R. Heathcote, 2005, ApJ 620, 321

\bibitem{zn}
H. Monteiro, H. E. Schwarz, R. Gruenwald, S. R. Heathcote, 2004, ApJ 609, 194
\bibitem{zo}
C. Morisset, G. Stasi\'{n}ska \& M.  Pe\~na, 2005, MNRAS 360, 499 

\bibitem{zp}
C. Morisset, D. P\'equignot, 1996, A\&A 312, 135

\bibitem{zq}
C. Mendoza, M. Seaton, S. Nahar, A. Pradhan, T.  Kallman, C. Zeippen, 2002, American Physical Society, 2002 Division of Atomic, Molecular and Optical, abstract J6.054

\bibitem{zr}
Nahar, S. N. 2005, ApJS 156, 93

\bibitem{zs}
Nahar, S. N. 2003, in \emph{Planetary Nebulae: Their Evolution and Role in the Universe}, Proc. of the 209th IAU Symp., S. Kwok, M. Dopita and R. Sutherland eds. ASP, p.335

\bibitem{zt}
R. Neal, B. Ercolano \& R. Sahai, MNRAS, in preparation

\bibitem{zu}
C. R. O' Dell, W. J. Henney \& G. J. Ferland 2005, AJ 130, 1720

\bibitem{zv}
S. Och, L. B. Lucy \& M. R. Rosa, 1998, A\&A 336, 301

\bibitem{zw}
D.~E. Osterbrock, in \emph{Astrophysics of Gaseous Nebulae and Active Galactic Nuclei}, Mill Valley, CA: Univ. Sci. Press

\bibitem{zy}
M. Peimbert, 1967, ApJ 150, 825

\bibitem{zz}
D. P\'{e}quignot, X.-W.  Liu,  M. J. Barlow, P. J. Storey, C. Morisset, 2003, in \emph{Planetary Nebulae: Their Evolution and Role in the Universe}, Proc. of the 209th IAU Symp., S. Kwok, M. Dopita and R. Sutherland eds. ASP, 347

\bibitem{zza}
D. P\'{e}quignot et al. 2001, See Ferland \& Savin 2001, p. 533

\bibitem{zzb}
D. P\'{e}quignot, G. Stasinska \& S.~M.~V. Aldrovandi, 1978, A\&A 63, 313

\bibitem{zzc}
R. H. Rubin, S. W. J. Colgan, M. R. Haas, S. Lord, \& J. P. Simpson, 1997, ApJ 479, 332

\bibitem{zzd}
R.~H. Rubin, 1968, ApJ 153, 761

\bibitem{zze}
R. Sahai, A. Dayal et al., 1999, AJ 118, 468

\bibitem{zzf}
A. K. Speck, M. Meixner, G. H. Jacoby, P. M. Knezek, PASP 115, 170

\bibitem{zzg}
P. J. Storey, B. Ercolano, \& N. R. Badnel, MNRAS in preparation

\bibitem{zzga}
R. S. Sutherland \& M. A. Dopita 1993, ApJS, 88, 253

\bibitem{zzh}
 Y. G. Tsamis, M J. Barlow, X.-W. Liu, P. J.  Storey \&  I. J. Danziger, 2004, MNRAS 353, 953

\bibitem{zzi}
A. G. G. M. Tielens \& D. Hollenbach 1985, ApJ 291, 722

\bibitem{zzl}
R. Tylenda, 2003, in \emph{Planetary Nebulae: Their Evolution and Role in the Universe}, Proc. of the 209th IAU Symp., S. Kwok, M. Dopita and R. Sutherland eds. ASP, 389

\bibitem{zzm}
P. A. M. van Hoof, J. C. Weingartner, P. G. Martin, K. Volk \& G. J. Ferland 2004, MNRAS 350, 1330

\bibitem{zzn}
J. C. Weingartner \& B. T. Draine 2001, ApJS, 134, 263, 

\bibitem{zzo}
R.~E. Williams, M. Livio, eds. 1995. \emph{The Analysis of Emission Lines, Space Telesc. Sci. Inst. Symp. Ser.} Cambridge, UK: Cambridge Univ. Press

\bibitem{zzp}
K. Wood, J. S. Mathis, B. Ercolano, MNRAS 348, 1337 

\bibitem{zzq}
K. Wood, J. S. Mathis, 2004, MNRAS 353, 1126

\end{thebibliography}

\IfFileExists{\jobname.bbl}{}
 {\typeout{}
  \typeout{******************************************}
  \typeout{** Please run "bibtex \jobname" to optain}
  \typeout{** the bibliography and then re-run LaTeX}
  \typeout{** twice to fix the references!}
  \typeout{******************************************}
  \typeout{}
 }

%%%%%%%%%%%%%%%%%%%%%%%%%%%%%%%%%%%%%%%%%%%
%% The following lines show an example how to produce a bibliography
%% without the help of the BibTeX program. This could be used instead
%% of the above.
%%%%%%%%%%%%%%%%%%%%%%%%%%%%%%%%%%%%%%%%%%%

\end{document}